\documentclass[12pt, preprint]{aastex}
\usepackage{natbib}

\begin{document}

\title{Star Spot Induced Radial Velocity Variability in LkCa 19}

\author{Marcos Huerta\altaffilmark{1,2} and Christopher M. Johns-Krull\altaffilmark{2}}
\affil{Physics \& Astronomy Department, Rice University, Houston, Texas 77005}

\author{L. Prato\altaffilmark{2}}
\affil{Lowell Observatory, Flagstaff, AZ 86001}

\author{Patrick Hartigan}
\affil{Physics \& Astronomy Department, Rice University, Houston, Texas 77005}

\author{D. T. Jaffe}
\affil{Department of Astronomy, University of Texas at Austin, Austin, Texas 78712}

\altaffiltext{1}{Now at  Department of Astronomy, University of Florida, Gainesville, Florida 32611}
\altaffiltext{2}{Visiting Astronomer, McDonald Observatory, which is operated by the University of Texas at Austin}

\begin{abstract}

We describe a new radial velocity survey of T Tauri stars and present the first results.  Our search is motivated by an interest in detecting massive young planets, as well as investigating the origin of the brown dwarf desert.  As part of this survey, we discovered large-amplitude, periodic, radial velocity variations in the spectrum of the weak line T Tauri star LkCa 19.  Using line bisector analysis and a new simulation of the effect of star spots on the photometric and radial velocity variability of T Tauri stars, we show that our measured radial velocities for LkCa19 are fully consistent with variations caused by the presence of large star spots on this rapidly rotating young star.  These results illustrate the level of activity-induced radial velocity noise associated with at least some very young stars.  This activity-induced noise will set lower limits on the mass of a companion detectable around LkCa 19, and similarly active young stars.

\end{abstract}

\section{Introduction}

The first successful detection produced by radial velocity surveys of a planet orbiting a star occured in 1995 when a near Jupiter-mass planet was found around the main sequence star 51 Pegasus \citep{1995Natur.378..355M}.  Subsequent observations verified the existence of the companion and refined its orbital parameters \citep{1995AAS...187.7004M,1997ApJ...481..926M}.  By 2007, over 200 planetary systems have been discovered via the radial velocity method and the statistics of various properties of the planets have been catalogued \citep{2006ApJ...646..505B}

Most radial velocity surveys have been conducted on main sequence stars typically older than 1 Gyr.  The use of the radial velocity method on younger stars is challenging because of increased activity, specifically in the form of signals caused by star spots.  Spots on the surface of the star rotate in and out of view of the observer, potentially creating spurious radial velocity signatures.

\citet{2004AJ....127.3579P} looked for planetary companions to stars in the Hyades cluster ($\sim$790 Myr).  They found activity induced jitter consistent with the empirical relationship between activity and radial velocity described in \citet{1997ApJ...485..319S}.  \citet{2004AJ....127.3579P} also found one binary ($K = 1152 $ m s$^{-1}$) and two candidate binaries with linear velocity trends, but no planetary companions.  More recently, \citet{2006PASP..118..706P} examined 61 stars in the $\beta$ Pic moving group ($\sim$12 Myr) and the Ursa Majoris association ($\sim$300 Myr) and found no Jupiter mass planets in short period orbits.  The authors found these results unsurprising, as planetary companions are rare around main sequence stars with the small orbital separations ($<$ 0.1 AU) that these surveys were able to probe.  They measured maximum radial velocity amplitudes of 30-100 m s$^{-1}$ for targets in Ursa Majoris and 250-600 m s$^{-1}$ for targets in $\beta$ Pic.

These searches were motivated, as is ours, by an interest in understanding the formation of both stellar and planetary systems.  By probing the youngest stars for planetary companions, we can learn at what stage planets form.  Since the time scale for the core accretion model \citep{1996Icar..124...62P} is longer than formation due to disk instability \citep{1997Sci...276.1836B}, a detection of a young planet around a T Tauri star should also provide clues to the formation mechanism itself.  We can also explore the level of stellar activity in young stars and how this can effect our ability to measure radial velocity variations, and therefore detect planets.  Recently, \citet{2007ApJ...660L.145S} have reported the detection of a planet around the 100 Myr star HD 70573.

An additional motivation to search for companions around young stars is to understand the origin of the so-called ``brown dwarf desert."  As more planetary companions were discovered in the last decade, and the completeness of the surveys increased, \citet{2000PASP..112..137M} noted the dearth of substellar objects with masses greater than 20 M$_{\rm Jup}$ in close orbits ($\lesssim$ 3 AU).  This ``brown dwarf desert" was further quantified by \citet{2006ApJ...640.1051G} who analyzed an unbiased sample of nearby sun-like stars to measure the rate of stellar, brown dwarf, and planetary companions.  They found that $\sim$16\% of their sample stars had close (P $<$ 5 yr) companions with masses greater than that of Jupiter.  Of these, 11\% $\pm$ 3\% were stellar, $<$ 1\% were brown dwarfs, and 5\%$\pm$ 2\% were giant planets, where a brown dwarf is defined as an object in the mass range between 13 M$_{\rm Jup}$ to 80 M$_{\rm Jup}$.

\citet{2002MNRAS.330L..11A} suggested that the kinematic viscosity of a sufficiently massive circumstellar disk could cause brown dwarfs in close orbits, interacting dynamically with the disks, to catastrophically migrate towards the host star and be lost to mergers.  They predict that young stars should have an order of magnitude more brown dwarfs in close orbits than main sequence stars, as the time scale for catastrophic migration is $\sim$ 1 Myr.  A search for substellar companions around T Tauri stars could confirm or refute this scenario, improving our understanding of the brown dwarf desert in older stars.

In this paper, we summarize our spectroscopic and photometric observational techniques for our sample of T Tauri stars and report a detection of periodic radial velocity variations in the weak line T Tauri star LkCa 19, for which we have a large number of observations.  LkCa 19 has a  $v \sin i$ = 18.6 km s$^{-1}$ \citep{1987AJ.....93..907H} and spectral type K0 \citep{1988cels.book.....H}.  LkCa 19, with its short rotation period of 2.24 days \citep{1993A&A...272..176B}, provides an excellent example of the effects of star spots on radial velocity measurements.  Using new simulations, we show that the radial velocity properties of LkCa 19 are well modeled by the presence of one large spot.

\section{Sample, Observations, Data Reduction}
\subsection{Sample}
\label{section:technique_rv}

Our total sample consists of 100 young stars in the Taurus star forming region and the Pleiades open cluster, with a V magnitude range from 9 to 15.   Approximately one third of the Taurus objects are classical T Tauri stars, with the remainder weak-line T Tauri stars.  Stars in the Pleiades and a subsample of our Taurus objects were observed as part of the Space Interferometry Mission (SIM) Planetquest precursor science program.  These observations will identify Jupiter-mass and greater spectroscopic companions; these systems will be excluded from the SIM target list.

Most targets were observed at least once per observing run, weather permitting.  A subset of targets were selected for observation on each night of every observing run.  Since rotation periods for T Tauri stars are typically 10 days or less, this was done to quantify the radial velocity variability of a sub-sample, and to understand the possible sources of error and velocity jitter such as star spots or other sources of activity (e.g. accretion).

\subsection{Observations : Spectroscopy}

In 2004, we began a radial velocity survey to search for planets and brown dwarfs around a sample of T Tauri stars using the Coud\'e echelle spectrograph  \citep{1995PASP..107..251T} on the 2.7m Harlan J. Smith telescope at McDonald Observatory.   The spectrograph yields $R = \frac{\Delta \lambda}{\lambda} = 60,000$ with a 1.2 arcsecond slit.   Observations at the 2.7m Harlan J. Smith telescope were taken in November of 2004, December of 2004 to January of 2005, November of 2005, and February 2006. The Julian dates of these observations are listed in Table 1.
Current high precision radial velocity surveys routinely achieve velocity precision of  $\sim$ 3 m s$^{-1}$ \citep[e.g.][]{1996PASP..108..500B}, usually by the use of an iodine cell to superimpose calibration absorption lines onto the stellar spectrum.   However, because we expect T Tauri stars to have intrinsic radial velocity noise considerably higher than this due to photospheric actvitiy such as spots \citep[e.g.][]{2004AJ....127.3579P, 2006PASP..118..706P}, we did not require such high precision measurements.  We make use of thorium-argon lamps taken before and after each object.  This method also avoids contaminating the spectra with iodine absorption lines.
\clearpage
\begin{deluxetable}{rrrc}
\label{table:lkca19_rvs}
\tablewidth{0 pt}
\tabletypesize{\small}
\tablecaption{LkCa19 Radial Velocities}
\tablehead{

 \colhead{} & \multicolumn{2}{c}{m/s}\\
\cline{2-3}\\

\colhead{JD}           & 
 \colhead{RV} &  \colhead{Error} & \colhead{S/N}
}
\startdata

2453367.8252	&	2193.6	&	190.5	&	81.6	\\
2453369.6507	&	1995.0	&	141.2	&	87.6	\\
2453370.8191	&	540.5	&	140.1	&	52.0	\\
2453371.6180	&	1381.4	&	147.9	&	76.1	\\
2453372.6210	&	1092.6	&	153.4	&	27.5	\\
2453373.7105	&	1095.8	&	175.4	&	77.6	\\
2453374.7654	&	1213.0	&	331.5	&	35.4	\\
2453375.6835	&	684.5	&	186.8	&	59.7	\\
2453693.8006	&	1780.0	&	168.4	&	54.2	\\
2453694.8526	&	1193.0	&	208.9	&	58.4	\\
2453695.7514	&	1160.2	&	202.6	&	71.3	\\
2453696.7825	&	1555.0	&	140.5	&	81.5	\\
2453697.7609	&	451.8	&	151.4	&	78.5	\\
2453769.5967	&	309.0	&	140.4	&	66.5	\\
2453769.7395	&	803.8	&	158.5	&	58.5	\\
2453770.5880	&	2041.2	&	195.8	&	60.8	\\
2453770.7327	&	1924.2	&	182.8	&	69.2	\\
2453771.6556	&	192.6	&	280.6	&	54.0	\\
2453771.8006	&	505.7	&	140.4	&	46.6	\\
2453772.5921	&	2052.0	&	163.2	&	57.8	\\
2453773.5887	&	105.4	&	144.7	&	64.5	\\
2453773.7526	&	0.0	&	121.0	&	57.7	\\
2453775.6523	&	743.8	&	148.4	&	57.7	\\
2453775.7981	&	307.7	&	164.7	&	63.2	\\

\enddata

\end{deluxetable}
\subsection{Observations : Photometry}

We obtained near-simultaneous differential CCD photometric observations of LkCa 19, beginning in December of 2004.  The purpose of these observations was to look for any concurrent photometric variability and to see if any signal in the photometry is also present in the radial velocity variations.  Photometric variability in weak line T Tauri stars such as LkCa 19 is believed to originate in dark star spots moving in and out of view as the star rotates.  The resulting false radial velocity modulation shows a quarter-phase offset between the photometric and RV light curves \citep{2001A&A...379..279Q}.  A substellar companion around a T Tauri star should not produce a photometric signature, assuming there is no planetary transit.  

Photometry was obtained using the prime focus corrector \citep{1995PhDT.........8C} on the 0.8 m telescope at McDonald observatory.  Our observations were taken through Bessel V, B, and R photometric filters.  The field of view for the images was 42.6 x 42.6 arcminutes.  Integration times ranged from 3 to 300 seconds.

\subsection{Data Reduction}

\subsubsection{Spectroscopy}

All data reduction, including wavelength calibration was performed using standard echelle spectral reduction routines in IRAF.   The raw spectra were bias subtracted and flat fielded using the normalized spectrum of an internal lamp.  Optimal extraction, to remove cosmic rays and improve signal, was used on all targets except for the bright radial velocity standards.  The dispersion solution was derived from the time weighted average of thorium-argon lamps taken both before and after each image.  Individual thorium-argon lamps had lines identified visually and calibrated using the NOAO spectral atlas database.  We used 7 to 10 hand-chosen lines per order across $\sim$ 15 orders of the total spectrum for the dispersion solution.  The final fit was done in both the dispersion and cross-dispersion axis, and has a typical RMS value of about 0.002 pixels, or ~4  m s$^{-1}$, which is a negligible source of error for this program.

The RVSAO package in IRAF was used to determine the radial velocities.  IDL was used to further analyze and visualize the radial velocity data.  The \texttt{xcsao} task in the RVSAO package was performed on the extracted and wavelength calibrated spectra to determine the radial velocities.  This routine is a fourier-based cross-correlation program, described in detail by \citet{1998PASP..110..934K}.  Cross-correlation is a powerful tool when coupled with large spectral ranges to determine precise radial velocities.  We performed cross-correlation using several echelle orders, with each order covering approximately 100 angstroms.  In order to avoid spectral type mismatching, we used the target star itself as the template for the cross correlation.  Therefore, our velocities are measured relative only to one observation epoch.  A spectrum with relatively high counts and minimal cosmic ray contamination was chosen to be the template.

\subsubsection{Photometry}

Photometric data reduction followed standard procedures using IRAF including bias correction, overscan correction, and flat fielding.  Stellar objects are identified with the \texttt{DAOPHOT} package within IRAF and then aperture photometry was used on all stars in the field. The error in the photometry was estimated by selecting a background star, or several, and looking at the standard deviation of that star over all observations.  Such stars should show no photometric variability, and any scatter in their measurements provides an easy estimate on the accuracy of the photometry.  This error is $\sim$0.01 magnitude.

\section{Analysis}

\subsection{Radial Velocity Uncertainties}

In order to gain a better understanding of the accuracy of our radial velocity technique, we observed six radial velocity standard stars, selecting objects from \citet{2002ApJS..141..503N}, \citet{1996PASP..108..500B}, and \citet{1999ApJ...526..890C}.  By measuring radial velocities of these targets each night of each observing run, we estimate the systematic errors and the accuracy of the radial velocity measurements with our instrumental setup.  We observed three bright standard stars, in the 3rd to 4th V magnitude range, in addition to three fainter objects around 6th to 7th V magnitude.  These standard stars were reduced and analyzed in the same way as our target stars, using one epoch of the standard star itself as the cross-correlation template.  Within a given run, we obtained very low radial velocity scatter, however there are systematic effects between runs that appear in the observations of our standard stars.
\begin{deluxetable}{rrrc}[t!]
\label{table:rv_standards}
\tablewidth{0 pt}
\tablecaption{Radial Velocity Standards Observed}
\tablehead{
\colhead{Star}           & 
 \colhead{N} &  \colhead{V Mag} & \colhead{Standard Deviation (m/s)\tablenotemark{a}}
}
\startdata
107 Psc & 20 & 5.2 & 100\\
HD 4628 & 20 & 5.75 & 129 \\
Tau Ceti & 19 & 3.50 & 184 \\
HD 65277 & 15 & 8.12 & 110 \\
HD 80367 & 19 & 8.16 & 107 \\
HD 88371 & 18 &  8.43 & 132\\

\enddata
\tablenotetext{a}{Includes systematic error seen in Figure \ref{fig:standards}.}

\end{deluxetable}

\begin{figure}[t!]
\begin{center}

\plotone{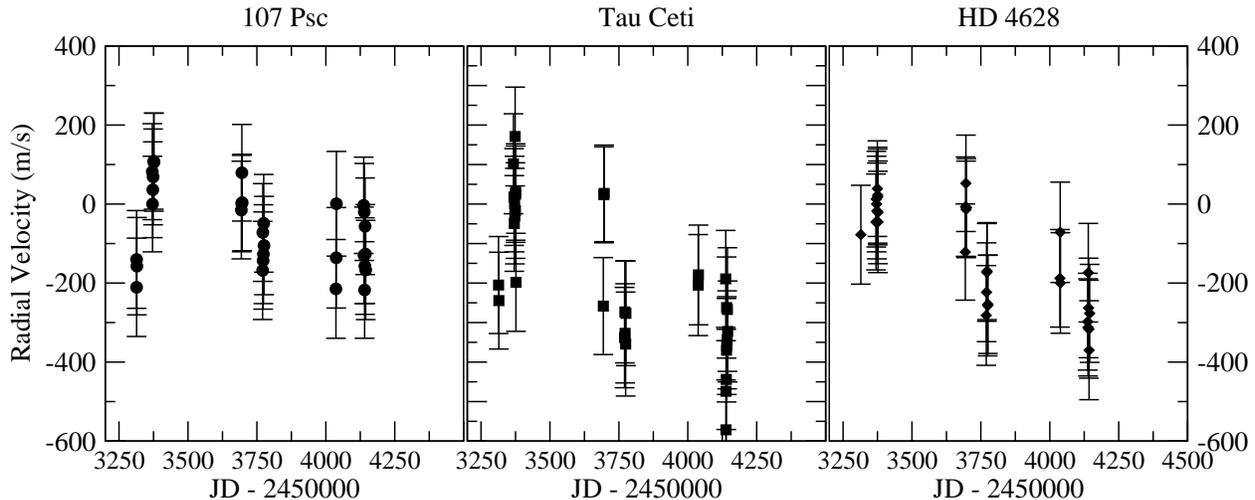}
\caption{Radial velocity shifts of the standards 107 Psc, Tau Ceti, and HD 4628 showing systematic offsets between observing runs.}
\label{fig:standards}
\end{center}
\end{figure}
Figure \ref{fig:standards} shows RV measurements for our 3 bright standard stars.  Each standard shows a distinct systematic offset between the several observing runs.  Simple explanations do not account for this systematic error -- the residuals in the dispersion solution look very similar in each run and we accounted for the motion of the earth itself.  We suspect the likely cause of the offset is the resetting of the slit plug from one observing run to the next, which could alter the position of the slit relative to the optical axis of the telescope and the thorium-argon lamp.  However, even with this systematic error uncorrected, the overall scatter of the standard stars remains $\sim$ 120 m s$^{-1}$ (see Table 2), which we believe is our nominal level of precision.  The final error listed in Table 1 is the sum in quadrature of this systematic error and error estimated from the order-to-order scatter in the measured radial velocities of LkCa 19.

\subsection{Detection of Periodic Signals}

Our T Tauri targets have considerably larger radial velocity variations, with standard deviations in the range of $\sim$ 200 - 3000 m s$^{-1}$, than the radial velocity standards.  LkCa 19 shows radial velocity variation with a standard deviation of $\sim$ 670 m s$^{-1}$, which is about six times higher than our nominal radial velocity standard.

\begin{figure}[t!]
\begin{center}
\plotone{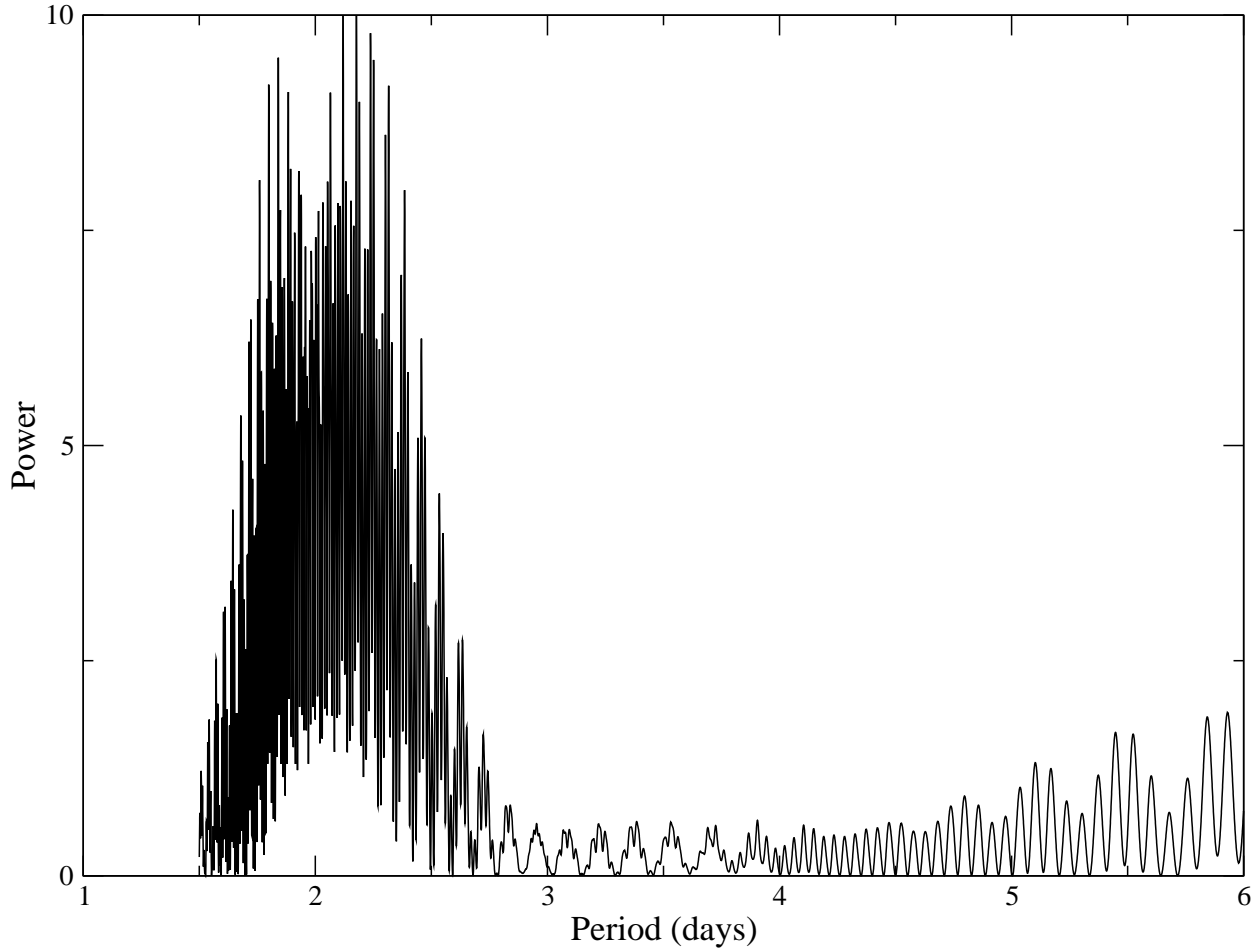}
\caption{The Scargle method power spectrum on all radial velocity points for LkCa19.  The peak with the largest power corresponds to a false alarm probability $< 0.001$ }
\label{fig:lkca19_power}
\end{center}
\end{figure}

\begin{figure}[t!]
\begin{center}
\plotone{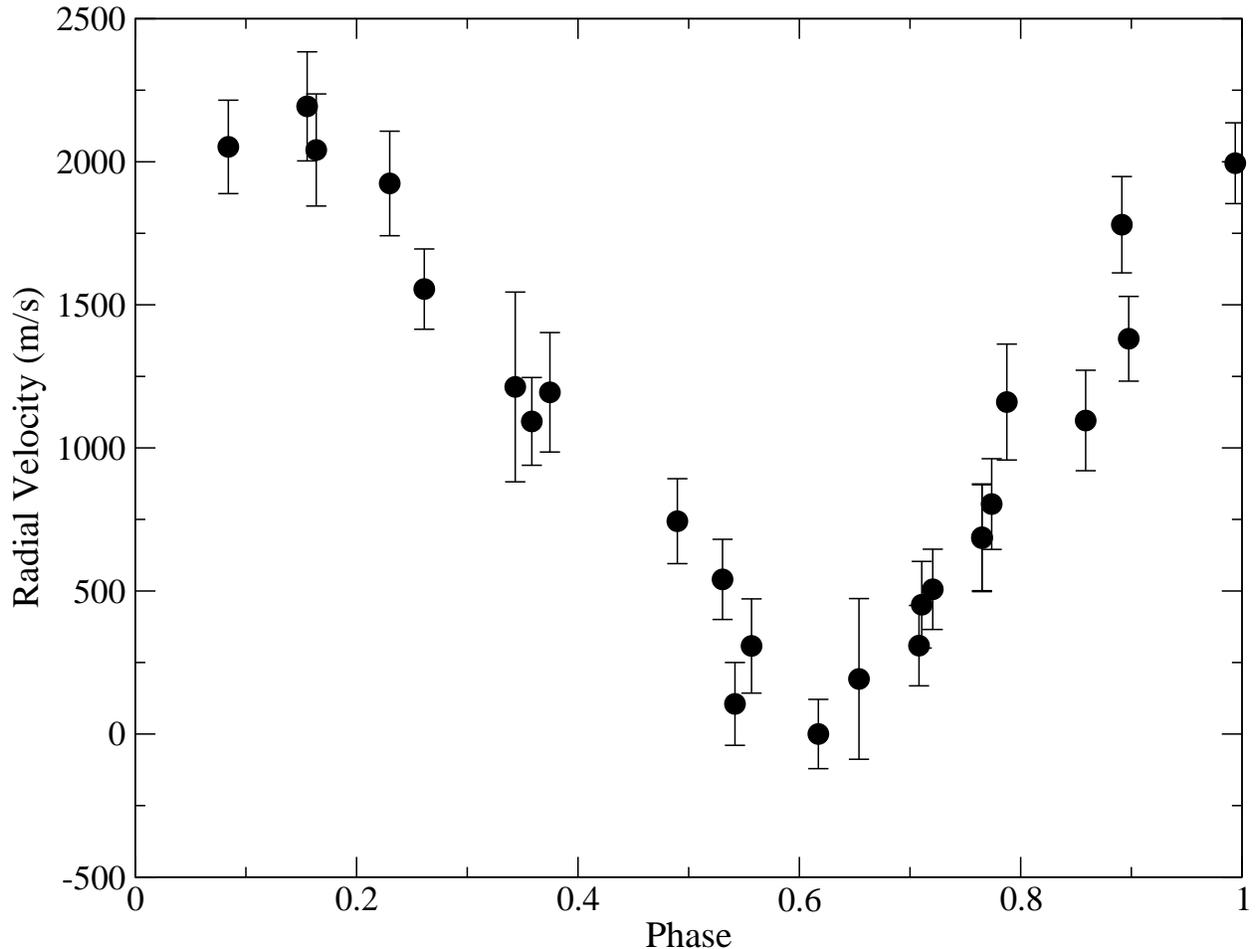}
\caption{Phase-folded LkCa 19 radial velocities with a period of 2.177 days.  The period was determined from a Scargle method power spectrum.  Error bars are a combination of the standard deviation of the mean of the six echelle orders used to compute radial velocity added in quadrature with the systematic errors from the radial velocity standards of 120 m/s.  No correction for the systematic errors between runs was applied.}
\label{fig:lkca19_phased_up}
\end{center}
\end{figure}

For targets with a significant number of observations, such as LkCa 19, we analyzed the data to look for any periodic variations of the radial velocity measurements.  This was done with an IDL implementation of the Scargle method of power spectrum estimation \citep{1982ApJ...263..835S}.  This method is best suited for our data, as it does not require the data to be evenly spaced in time.  The LkCa19 power spectrum is shown in Figure \ref{fig:lkca19_power}.
\clearpage
Once the initial power spectrum was computed over a wide range of periods, the power spectrum was computed again with dense sampling in order to resolve the strongest peak and to determine the best period, which we found to be 2.177 days with an uncertainty of 0.0264 days.  We then used the period to phase fold the original data so that any radial velocity periodicity could be evaluated (Figure \ref{fig:lkca19_phased_up}).

Further testing of the power spectrum analysis is necessary to ascertain the true significance of the detection of a period.  This was accomplished with a Monte Carlo simulation of our radial velocity data.  For LkCa19, 100,000 simulated sets of normally distributed random radial velocities with identical overall velocity scatter to our observed data were analyzed.  These random data sets were analyzed using the same temporal cadence as our actual observations.  None of these instances of random velocity data had a peak as strong or stronger in its power spectrum analysis than is present in the actual LkCa 19 data.  This indicates a high significance of this period detection, and a false alarm probability (FAP) of less than 10$^{-5}$.  Thus, we conclude that LkCa19 shows periodic radial velocity variations with a period of 2.177$\pm$0.0264 days.  However, this period is very near the 2.24 day rotation period of LkCa 19 \citep{1993A&A...272..176B}, suggesting that the observed signal is not the result of orbital motion.  Additional periods in the power function occur near 2.238 and 2.251 days, which are near the power of the strongest peak at 2.177 days.  This is an additional indication that we our observing the rotation period of the star.  The application of a correction for the systematic effects seen in Figure 1 produced no noticeable change in the Scargle periodogram of LkCa19, and no change in any of the detected periods.

\subsection{Bisector Analysis}

Bisector analysis of line profiles was first suggested by \citet[][]{1976oasp.book.....G} as a means to quantify the asymmetry of a specific absorption or emission line. Gray's interest was in the intrinsic line profile shapes of stars of various spectral types, and as a probe of granulation due to convection.  Bisector analysis as a means to check on the origin of radial velocity variations started with analysis of the original extrasolar planet 51 Pegasus \citep{1997ApJ...478..374H}.  The method was used subsequently by \citet{2001A&A...379..279Q} to look for a correlation between the slope of the bisectors of the spectrum's cross correlation function and the measured radial velocity of the star.  Similar analysis was done by \citet{2005A&A...442..775M} on several potential extrasolar planets, and by \citet{2007A&A...463.1017B} to test the origin of apparent radial velocity variations in the T Tauri star AA Tau. We performed bisector analysis on the cross correlation function (CCF) used to measure the radial velocity shift from one observation to another to test whether spots might be the cause of the observed radial velocity variations in LkCa 19.

\begin{figure}[t!]
\begin{center}
\plottwo{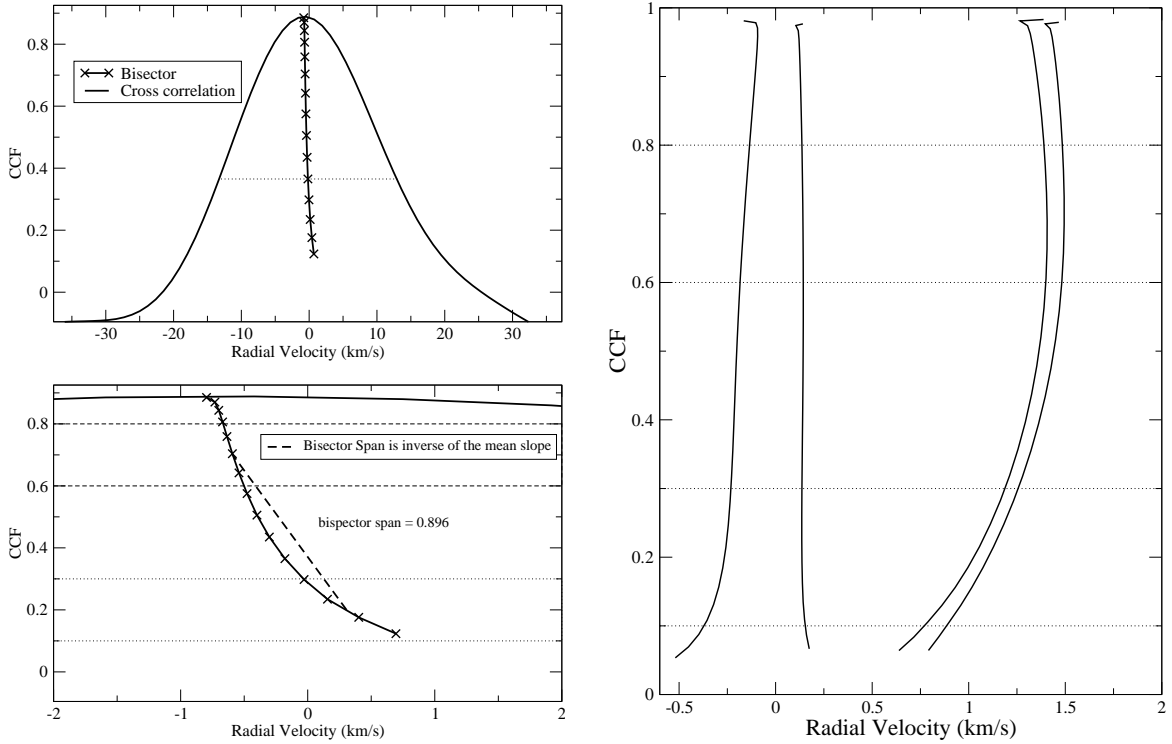}{f4b}
\caption{Top Left:  A cross-correlation function and its bisector.  The dotted line connects opposite sides of the CCF curve - the midpoint of this line represents one of the points of the bisector.  Bottom Left: Two mean velocities are computed - one in the upper region denoted by the dashed lines and another between the dotted lines.  The bisector span is the difference between these two velocities.  Right:  Four bisectors for LkCa19 at opposite phases of the radial velocity curve.  The changing asymmetry of the profile is noticeable.  Dotted lines indicate the CCF power ranges in which the span is measured.}
\label{fig:what_is_a_bisector}
\end{center}
\end{figure}

\begin{figure}[t!]
\begin{center}
\plottwo{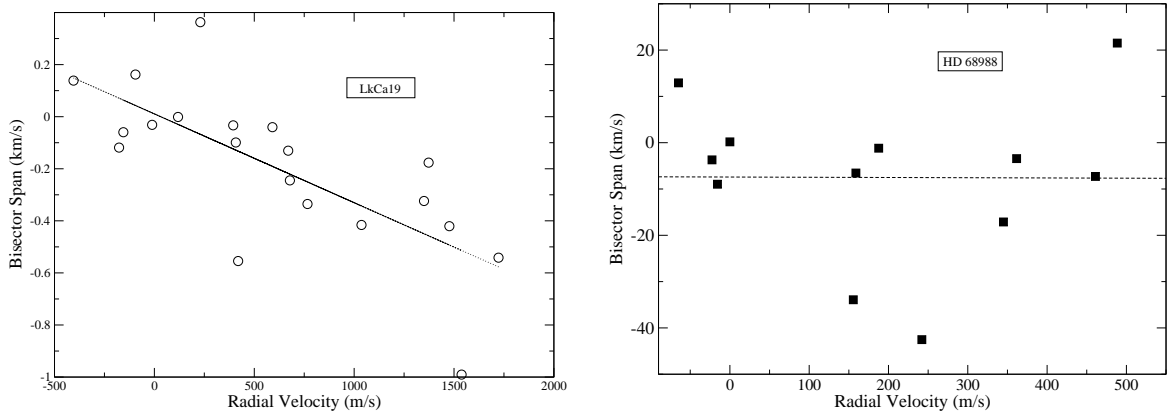}{f5b}
\caption{Left: LkCa19 bisector analysis.  The radial velocity and bisector appear to be correlated, with a linear correlation coefficient of $-$0.73.  The slope is consistent with the spot induced radial velocity variations of \citet{2001A&A...379..279Q}.  Right: HD 68988 bisector analysis.  No correlation is found.}
\label{fig:lkca19_bisector}
\end{center}
\end{figure}

To verify that we do not see correlated radial velocity and CCF bisector changes in systems with companions, we analyzed our data on a known extra solar planet host from the Keck High Precision Survey \citep{2002ApJ...568..352V}.  HD 68988 was observed to test our ability to measure a low amplitude periodic signal.  The ``span" of the bisector is similar to the slope of a line; it is measured by comparing the values of the bisector at two distinct heights of the cross correlation function (Figure \ref{fig:what_is_a_bisector}).  Bisector analysis of HD 68988 yielded no correlation between the bisector span and radial velocity, consistent with its classification as a planet-hosting star.  However, analysis of the bisectors for LkCa 19 through the most recent observing runs indicate a strong correlation between the bisector span and the radial velocity, a clear indication of star spot-induced radial velocity variations.  These results are shown in Figure \ref{fig:lkca19_bisector}.

\begin{figure}[t!]
\begin{center}
\plotone{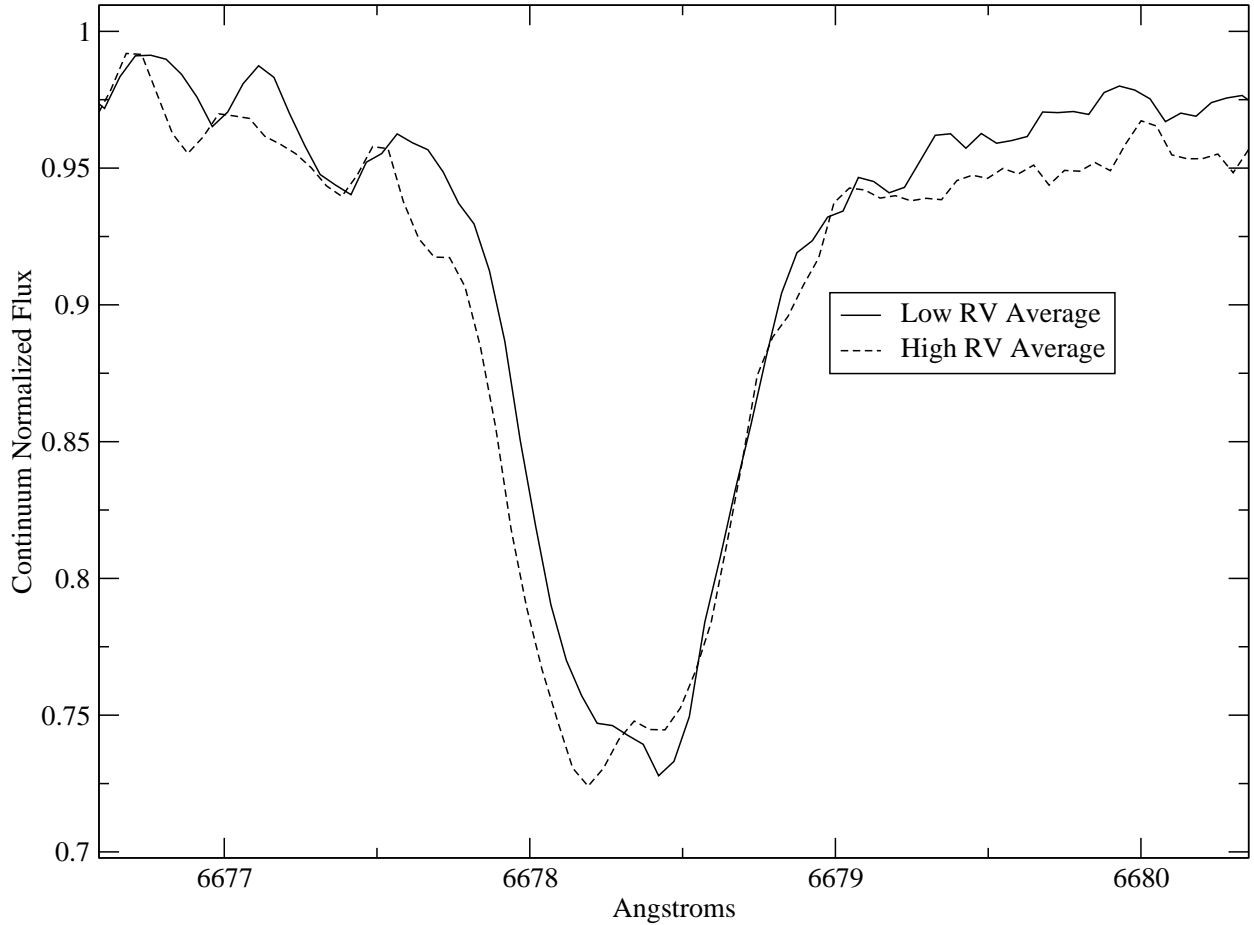}
\caption{LkCa 19 line profiles of the He I line at 6678 \AA{}.  The ``high" and ``low" radial velocity averages were created by averaging observations of similar radial velocity, 3 spectra from the high velocity case and 4 from the low velocity case, from consecutive nights of observing run in February of 2006.  The spectra were heliocentric corrected before averaging and smoothed with a moving average of 3 pixels after they were combined.}
\label{fig:lkca19_spectra}
\end{center}
\end{figure}

It is possible to see the variation in line profile shape in the LkCa 19 spectra.  In order to maximize signal to noise, we co-added several nights of similar velocity from our observations in February of 2006.  In addition, the averaged spectra were continuum normalized and smoothed with a moving average of 3 pixels.  The shifting asymmetry of these line averages can be seen in Figure \ref{fig:lkca19_spectra}.

\subsection{Photometric Analysis}

Photometric observations were obtained nearly simultaneously with our spectra of LkCa 19 to further aid the interpretation of any radial velocity variations.  The photometric light curves and any periodicity can be compared to the radial velocity curves and detected periods.

A FORTRAN code provided by J. Anderson was used to analyze the multiple exposures over many nights and to create light curves for our target stars.  The photometric zero-point was determined using a sigma-clipped average of all the stars in the field.  The photometry was phase-folded using the radial velocity period found in \S 3.2.  Furthermore, the photometry was analyzed to look for other periodicity.

\begin{figure}[t!]
\begin{center}
\plotone{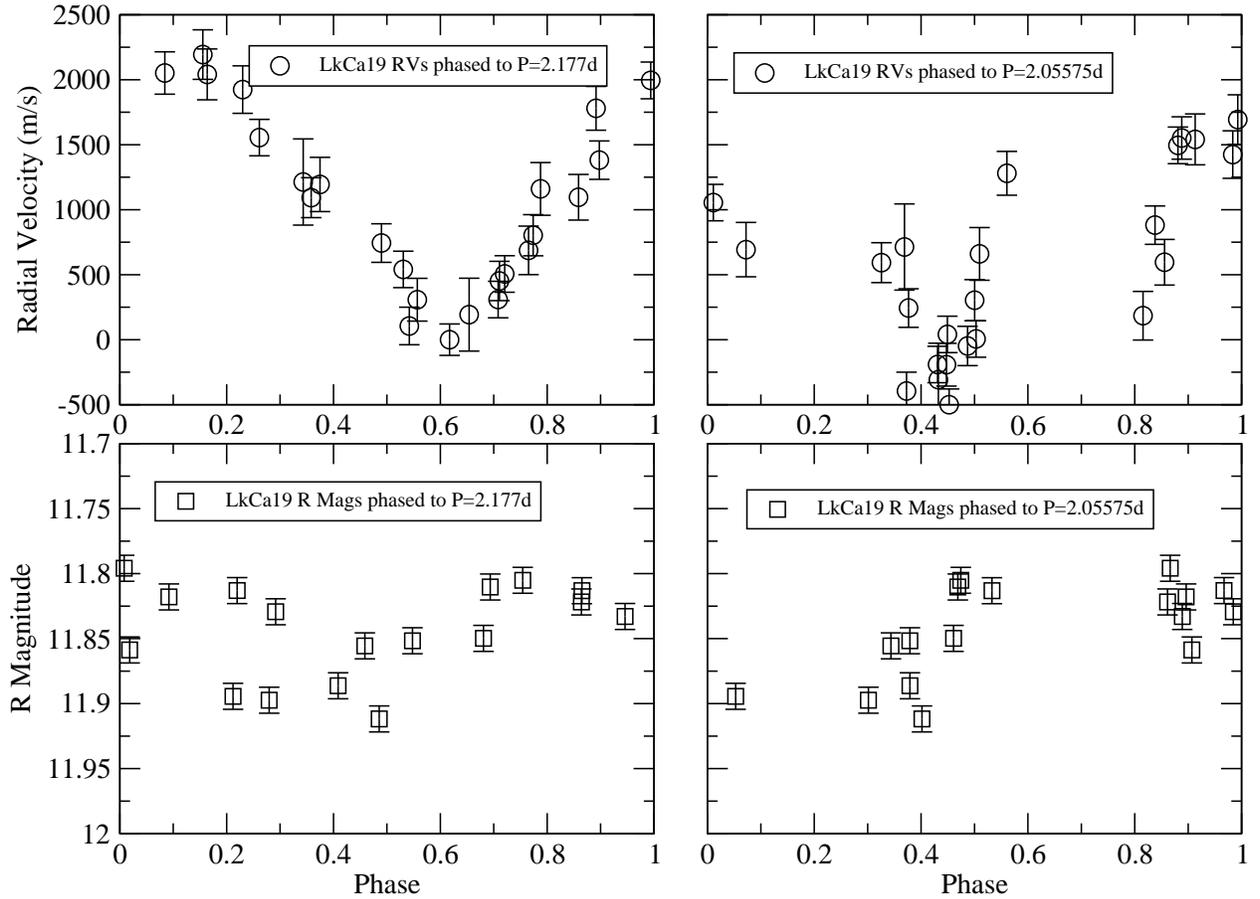}
\caption{LkCa19 RV and R-band photometry phased to periods of 2.177 and 2.05575 days.}
\label{fig:lkca19_phased_both}
\end{center}
\end{figure}

We analyzed 41 R band photometric points for LkCa 19, which are binned into nightly averages producing 17 photometric measurements.  On any given night, the R-band images were taken within $\sim$20 minutes of each other.  The data points are plotted, phased to the radial velocity period of LkCa 19, in Figure \ref{fig:lkca19_phased_both}.  The photometry does not appear in phase with the radial velocity signal.  A Scargle power spectra analysis for our photometry yields a period of 2.05575 days.  However, this photometric period determination does not withstand Monte Carlo test scrutiny, and the false alarm probability for the 2.05575 day LkCa 19 photometric period is 40\%.  The poor strength of this period is likely due to the smaller number of measurements compared to the radial velocity data, and the greater noise in the photometric measurements compared to the amplitude of the photometric variability.  This period, however, is close to both the detected radial velocity period and the known 2.24d rotation period of LkCa 19 \citep{1993A&A...272..176B}.  The peak-to-peak amplitude in the R-band is 0.12 magnitudes.  We observed the same amplitude in V-band, however in the B-band the peak-to-peak amplitude is larger at 0.18 magnitudes.

\section{Star Spot Models}
\subsection{Conceptual Framework}

The goal of our star spot modeling was to test the hypothesis that at least some of the radial velocity signatures observed in T Tauri stars can be attributed to the star's rotation combined with the presence of low temperature spots on the surface of the star.  In order to test these assumptions, it is necessary to numerically replicate the effect of star spots on a synthesized stellar spectrum.

To accomplish this task, a stellar disk integration model was constructed.  The apparent stellar disk is divided into sections, allowing for portions to be at a lower temperature and therefore lower intensity than the rest of the star, with appropriately different spectral features.  With the modeled stellar disk divided between a stellar spectrum at one effective temperature and one or more localized spots with a lower temperature, an integrated spotted star spectrum can be created and analyzed in the same manner as the observed stellar data.  This modeling involved integrating an artificial stellar disk, with control over the location and size of spots on the apparent disk, and was implemented in the IDL language.

In order to construct the apparent stellar disk of a star containing spots, it was necessary to first synthesize appropriate spectra for both the star and the spot.  The input spectra were produced using the SME (Spectroscopy Made Easy) code of \citet{1996A&AS..118..595V}.  We chose a 100 \AA{} bandpass near 6300 \AA{}, similar to one of the echelle orders in our actual data.  The SME synthesis code utilized NextGen atmospheric models \citep{1995ApJ...445..433A} and a line list from the Vienna Atomic Line Data (VALD) database.   The dispersion of the artificial spectra was 0.02 \AA{} per pixel, with the input and output spectra having 5000 pixels.  The stellar temperature was chosen to be 4300K and the spot temperature 2900K, based on observations and modeling of V410 Tau \citep{1994A&AS..107....9P}.  This produced a brightness ratio between the spots and the stellar surface of 0.08.

Spots were defined by the colatitude and longitude of the spot center, and the radius of the spot.  The spot's projected shape and location on the stellar disk were used to trigger the integration routine to use the lower temperature and intensity spot spectrum rather than the standard stellar spectrum when the given portion of  the stellar disk is included.  Each grid point of the star that was deemed to be inside the spot was assigned the cooler, dimmer, spot spectrum.  Additionally,  each grid point has the corresponding radial velocity shifts based on the rotation of the star.  Therefore, Doppler broadening of the line profiles was handled within the disk integration itself.  Stellar limb darkening was handled by using the appropriate SME spectra for multiple limb angles.  The disk integration's output was tested against the disk integration output of the SME code -- a star with no spots -- and correctly returned a spectrum identical to the output of the original SME disk integration.

\subsection{Analysis of Simulated Spectra}

The simulation produced a series of spectra, which were then read into IRAF and analyzed for radial velocities in a manner identical to the actual observations.  Because the relative spectral fluxes are preserved, the IRAF bandpass spectrophotometry task \texttt{sbands} was used on a continuum region of the spectrum to estimate the photometric variations of the star due to the spots, using a wavelength bandpass of 1 \AA{} wide centered at 6235.5 \AA{}.

Twenty synthesized spectra were created representing 20 evenly spaced rotation steps of 18 degrees and thus one complete stellar rotation.  Each spectrum was cross-correlated against a template synthetic spectrum of the same $v \sin i$  but without any star spots.  The primary difference in the model analysis was that, without multiple echelle orders, we could not estimate the radial velocity error from the order to order scatter.  The artificial spectra were otherwise run through identical analysis as the actual data, although they were not rebinned or resampled to the resolution of the observations, nor was any noise added.  The RVSAO package was used to compute the velocities of the simulated spectra, and the bisector analysis was performed in the same manner as for the actual stellar data. 
 
The simulation code can be used with a variety of stellar spot configurations, ranging from a simple equatorial spot in an uninclined star, to a polar spot on a star with a tilt, to a multispot configuration, or any combination of the above.  This wide range of possibilities led to the question of which scenarios to simulate, and if any of them could reproduce the magnitude of radial velocity variations seen in our program stars.  We discuss three specific cases.  All of the computations were done with a $v \sin i$  value of 20 km/s, which is near the value for LkCa19, unless otherwise noted.

 \subsubsection{A Circumpolar Spot}
 \label{sec:circumpolar}
\begin{figure}[t!]
\begin{center}
\includegraphics[angle=90,scale=.80]{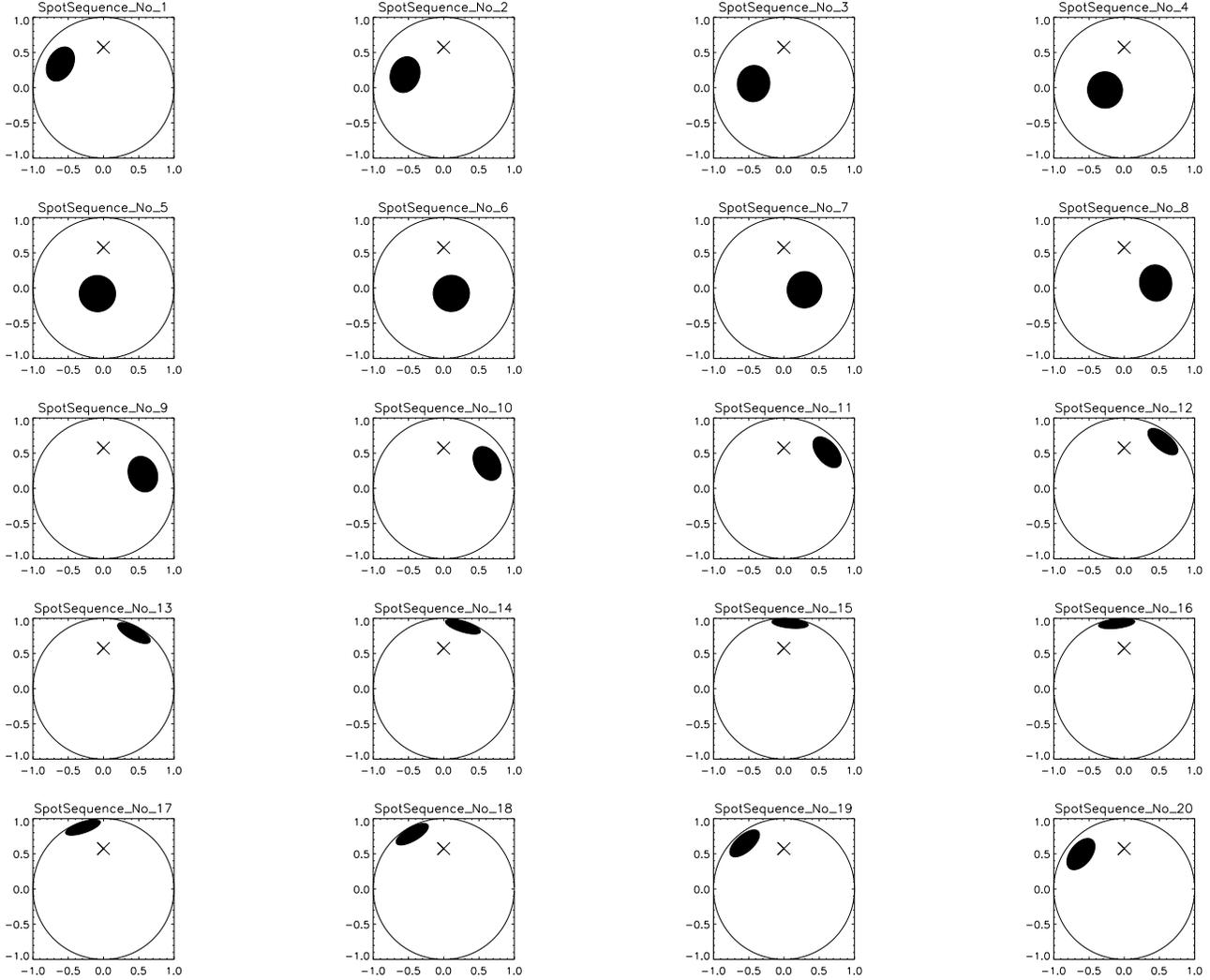}
\caption{The rotation sequence of a single spot rotating around the pole of an inclined star; i =  35$^\circ$ and the spot's colatitude is 40$^\circ$.   The X marks the pole.}
\label{fig:rotation_sequence_onepolarishspot}
\end{center}
\end{figure}

A single equatorial spot would yield a flat radial velocity curve while the spot is on the opposite side of the star from the observer.  In order to produce, with a single spot, a radial velocity curve that shows persistent amplitude modulation over time, it is necessary for the spot to always be visible to the observer.  Therefore, a spot close to the pole of the star, with the star inclined towards the observer, has been chosen as a simple geometry to model.  In this case, the spot has a colatitude of 40$^\circ$ and a radius of 15$^\circ$; the stellar inclination is 35$^\circ$.  Several Doppler imaging surveys have found polar and near polar spots \citep[e.g.][]{1994A&A...285L..17S,1994A&A...291L..19J,1997ApJ...487..896J,2004MNRAS.348.1301U}.  The circumpolar spot rotation sequence is illustrated in Figure \ref{fig:rotation_sequence_onepolarishspot}.

\begin{figure}[t!]
\begin{center}
\epsscale{1}\plottwo{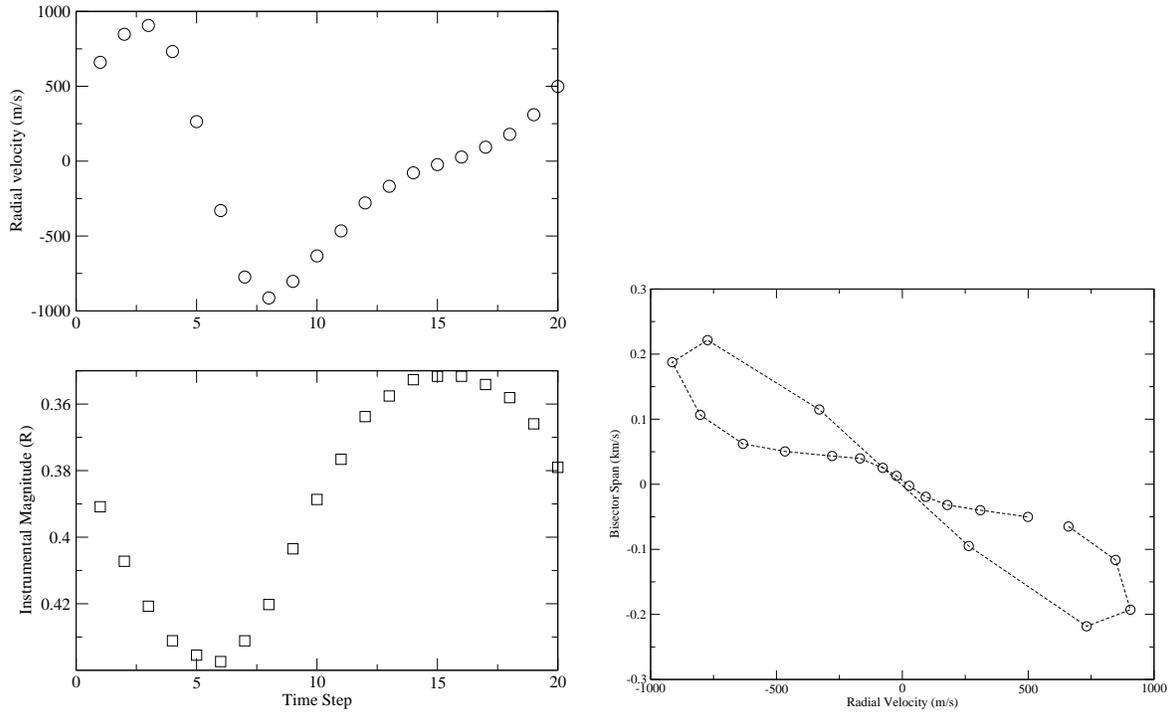}{f9b}
\caption{Radial Velocity and Photometric curves for the single circumpolar spot (left), and the corresponding bisector span vs. radial velocity plot (right).}
\label{fig:rv_and_phot_curves_onepolarishspot}
\end{center}
\end{figure}

The photometric light curve and the radial velocity curve are shown in Figure \ref{fig:rv_and_phot_curves_onepolarishspot}.  This spot configuration produces a radial velocity curve that varies continuously and has an equally regular photometric curve.  For this spot the photometric variation is at most 0.086 magnitudes and the full radial velocity amplitude is 1820 m/s.  We adopt a maximum flux-weighted filling factor for the spot following \citet{1996ApJ...463..766O}, who define the flux weighted filling factor as the total fractional projected area of spots on the observed hemisphere weighted by limb drakening.  This is readily computed via our disk integration code and is 9.1\% compared to the predicted values of 8.7\% of \citet{1997ApJ...485..319S} and 6.4\% of \citet{2002AN....323..392H}.  Both papers contain an empirical relationship between the related filling factor, radial velocity amplitude, and  $v \sin i$.

Bisector analysis of this spot configuration (Figure \ref{fig:rv_and_phot_curves_onepolarishspot}) shows the same general trend as the bisector analysis of LkCa 19. The width of the linear bisector span - velocity correlation appears intrinsic, not merely an artifact of noise in the actual stellar data.  The correlation between the span and the measured velocity is degenerate, in that the same velocity does not always produce an identical bisector span, even for one unique spot configuration. 

\subsubsection{Fifteen Random Spots}

\begin{figure}[t!]
\begin{center}
\includegraphics[angle=90,scale=.80]{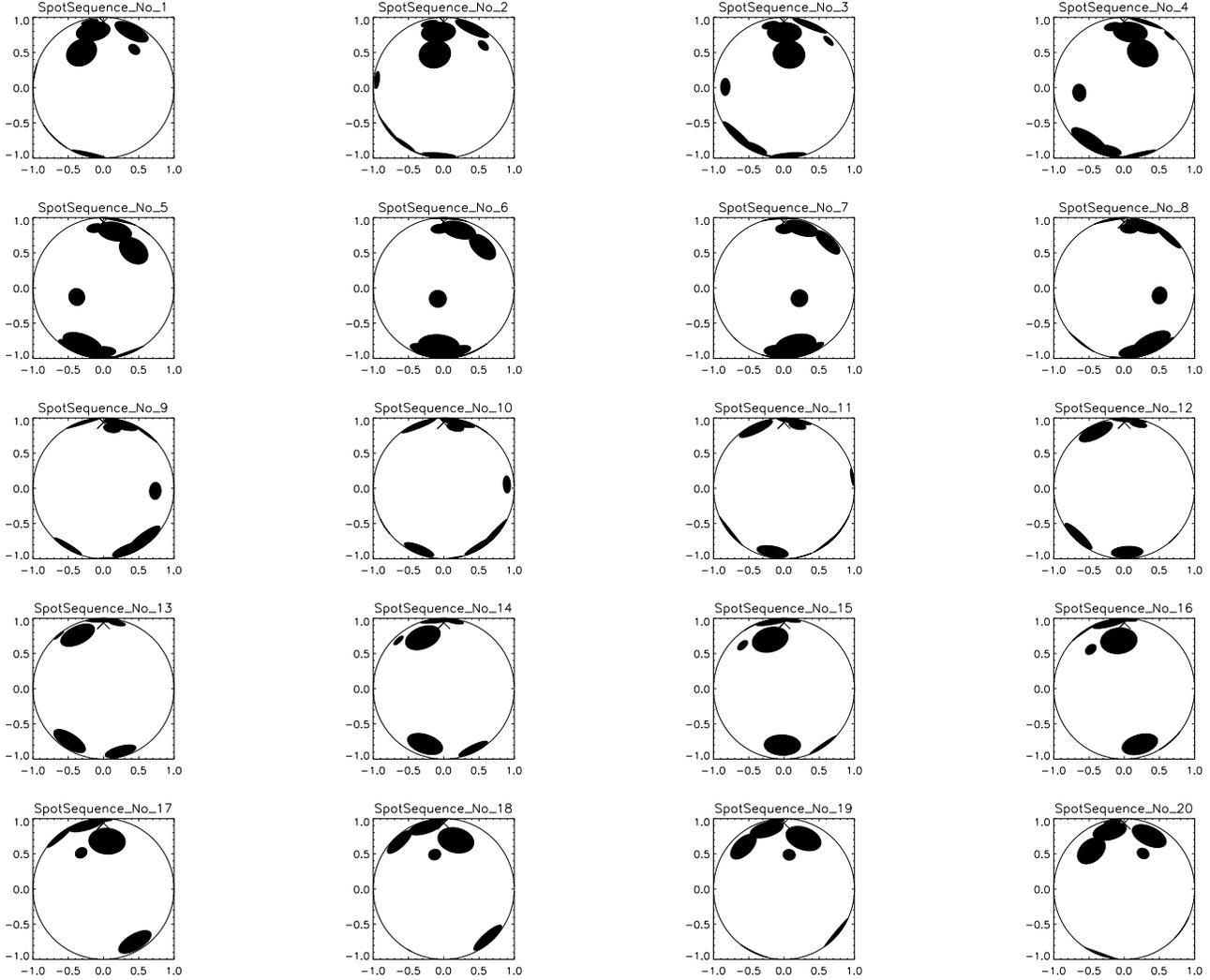}
\caption{The rotation sequence of fifteen random spots.  i =  86$^\circ$.}
\label{fig:rotation_sequence_fifteen_random_spots1}
\end{center}
\end{figure}

In addition to a simple one spot model, we performed a simulation that involves many spots, randomly distributed, with the star at a random inclination.  The random spot distribution was accomplished using the \texttt{randomu} task in IDL to produce 15 uniformly randomly distributed colatitude and longitude pairs.  The sizes of the spots were fixed with five spots each of three different sizes (Figure \ref{fig:rotation_sequence_fifteen_random_spots1}).
\clearpage

\begin{figure}[t!]
\begin{center}
\epsscale{1}\plottwo{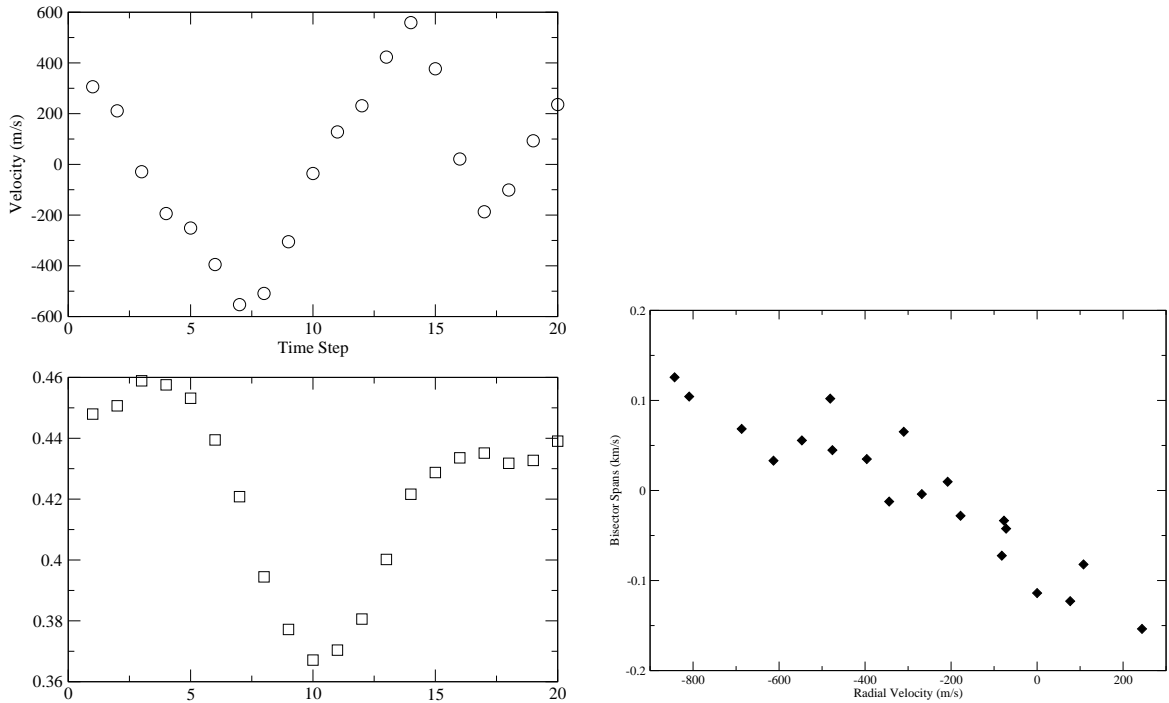}{f11b}
\caption{Radial velocity and photometric curves for fifteen random spots (left), and the corresponding bisector span vs. radial velocity (right).  The correlation coefficient of a linear regression fit is -0.91, indicating a strong correlation.}
\label{fig:rv_and_phot_curves_fifteen_spots}
\end{center}
\end{figure}

Models with multiple spots of different sizes produced very different radial velocity and photometric light curves (Figure \ref{fig:rv_and_phot_curves_fifteen_spots}) than the simpler one spot model.  The bisector analysis demonstrates the intrinsic scatter of the bisector span - radial velocity correlation.  There is a degeneracy in the bisector correlation such that the bisector alone can not predict the measured velocity.  The trend observed in this simulation is consistent with the correlation seen in LkCa 19.  The most significant feature of this simulation is the amplitude of the radial velocity curve, roughly 1200 m/s, and the amplitude of the photometric measurements, $\sim$ 0.1 magnitudes.

\begin{deluxetable}{llll}

\tablewidth{0 pt}
\tablecaption{Correlation Statistics for Bisector Span and Radial Velocity}
\tablehead{
\colhead{Object}           & 
 \colhead{Spearman $\rho$\tablenotemark{a}} & \colhead{Two Sided Significance} & \colhead{Linear R}
}
\startdata
LkCa19 & -0.79 & $3 \times 10^{-5}$ & -0.73 \\
Simulation (15 Spots) &-0.94 & $1.4 \times 10^{-9}$ & -0.93 \\
Simulation (One Circumpolar Spot) &-0.96 & $8.6 \times 10^{-12}$ & -0.92 \\

\enddata
\label{table:bisector_correlation}

\tablenotetext{a}{Rank Statistic}
\end{deluxetable}

\subsection{The Relationship between Photometric and Radial Velocity Amplitude}
\label{sec:phot_rv_amplitude}
\begin{figure}[t!]
\begin{center}
\epsscale{.7}\plotone{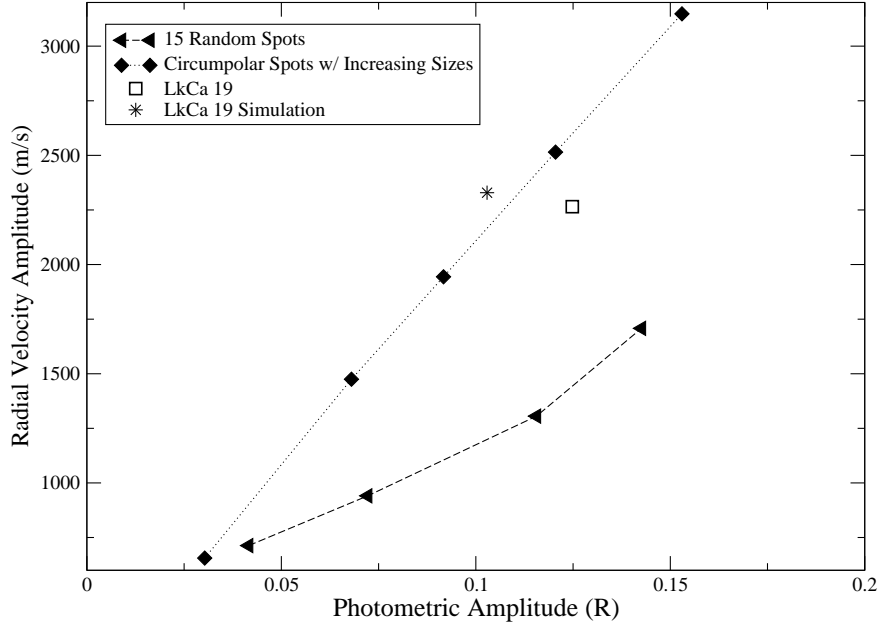}
\caption{Full Radial Velocity amplitude versus full photometric amplitude for a variety of simulations, observations of LkCa 19, and our simulation of LkCa 19.}
\label{fig:amplitudes_rv_and_phot}
\end{center}
\end{figure}

The results from the three example spot simulations, along with other spot geometries not described here in detail, prompted an inquiry into the relationship between the photometric amplitude and the radial velocity amplitude.  We compared the photometric and radial velocity amplitudes of several simulations, along with our actual observations.  Included were four circumpolar spots of various sizes similar to the spot sequence described in \S\ref{sec:circumpolar}.  Similarly, several additional spot simulations that contained 15 random spots were run, each with a different set of spot locations, but with the same inclination and spot size distribution.  The results for the simulations and LkCa19 are plotted in Figure \ref{fig:amplitudes_rv_and_phot}.  In all cases $v \sin i$ = 20 km s$^{-1}$.

LkCa 19 falls very close to the line representing circumpolar spots of various sizes.  The multi-spot configuration has a distinctly different slope.   The LkCa 19 simulation point in Figure \ref{fig:amplitudes_rv_and_phot} represents a best guess at the parameters of LkCa 19 itself.  The inclination angle was estimated at 25$^\circ$ based on the measured $v \sin i$ value \citep{1988cels.book.....H} and an inferred radius based on its luminosity and temperature.  The actual $v \sin i$  value of 18.6 km/s was also used in this simulation, otherwise the parameters are similar the circumpolar spot model described earlier.

\section{Conclusion}

Analysis of early results from our radial velocity survey of young stars reveals large radial velocity variations in our target stars compared to radial velocity standards.  LkCa 19 shows a radial velocity signal with a significant periodic variation.  Initial analysis of the radial velocities of LkCa 19 led to twice-nightly observations in our February 2006 run to obtain better phase coverage.  Bisector analysis, however, indicates that the radial velocity variations in LkCa 19 are very likely induced by star spots, as does the similarity between LkCa19's 2.177$\pm$0.0264 day radial velocity period and its known rotation period of 2.24 days \citep{1993A&A...272..176B}.  

While it is possible that a close planet could experience tidal locking \citep{1997ApJ...481..926M} and thus have an orbital period identical to the rotation period of the star, it is more likely we are seeing radial velocity modulation arising in star spots.  A surprising feature of LkCa 19 is the phase coherence of the spot-induced radial velocity modulation.  The velocities remain in phase at the period of 2.177 days over a time span of approximately 13 months.  This result is similar to the false planet detection reported by \citet{2001A&A...379..279Q}, also attributed to star spots, which showed reasonable phase-coherence for a time span of roughly 2 years.

We have also presented a new simulation of the effects of a rotating star with spots on line profiles and have measured the radial velocities and photometric variations from spectra produced by the simulation.  The models can reproduce the photometric and velocity amplitudes of LkCa 19.  All simulations show a strong and significant bisector correlation, summarized in Table 3. Further, each spot simulation shows a 0.25 lag in phase between photometry and radial velocity, regardless of spot configuration, which is expected for cool spots.

The spot geometry with the most regular or planet-like radial velocity curve is that of a single spot near the pole of a star with a low inclination angle.  A highly spotted star with the spots set randomly on the stellar surface, produces much more irregular radial velocity and photometric light curves.

If the magnitude of the photospheric activity induced radial velocity variations of LkCa 19 is typical of young stars, then our ability to find companions to these objects will be limited to the extremely massive and short-period sub-stellar companions.  LkCa19 has an exceptionally short rotation period and relatively high $v \sin i$  value, which offers some hope that its spot-induced radial velocity variations are not necessarily typical for T Tauri stars. 

\acknowledgments

We made use of IDL libraries available from the NASA Goddard Space Flight Center, and the John Hopkins / Advanced Physics Lab / Space Oceanography group, as well as the IRAF, originally from the National Optical Astronomy Observatory, and now maintained by the volunteers at IRAF.net.  Jay Anderson provided very helpful assistance with the photometric reduction, and useful discussions regarding the spot simulation code.  The authors are grateful to the anonymous referee for useful comments. This research also made use of NASA's Astrophysics Data System.  M.H. was partially supported by NSF Cooperative Agreement Number HRD-0450363 as a part of the Rice University Alliance for Graduate Education and the Professoriate (AGEP) Program.  Additional support was provided through NASA Grant 05-SSO05-0086 to Lowell Observatory (PI -- L. Prato), and from the Space Interferometry Mission Key Project, ``A Search for Young Planetary Systems and the Evolution of Young Stars" (PI -- C. Beichman).

\bibliography{huerta}

\end{document}